# Probing DNA interactions with proteins using a single-molecule toolbox: inside the cell, in a test tube, and in a computer

Adam J. M. Wollman, Helen Miller, Zhaokun Zhou, Mark C. Leake*

*Biological Physical Sciences Institute (BPSI), Departments of Physics and Biology, University of York, York YO10 5DD, UK.*

*mark.leake@york.ac.uk

## Abstract

DNA-interacting proteins have roles multiple processes, many operating as molecular machines which undergo dynamic metastable transitions to bring about their biological function. To fully understand this molecular heterogeneity, DNA and the proteins that bind to it must ideally be interrogated at a single molecule level in their native *in vivo* environments, in a time-resolved manner fast to sample the molecular transitions across the free energy landscape. Progress has been made over the past decade in utilising cutting-edge tools of the physical sciences to address challenging biological questions concerning the function and modes of action of several different proteins which bind to DNA. These physiologically relevant assays are technically challenging, but can be complemented by powerful and often more tractable *in vitro* experiments which confer advantages of the chemical environment with enhanced detection single-to-noise of molecular signatures and transition events. Here, we discuss a range of techniques we have developed to monitor DNA-protein interactions *in vivo*, *in vitro* and *in silico*. These include bespoke single-molecule fluorescence microscopy techniques to elucidate the architecture and dynamics of the bacterial replisome and the structural maintenance of bacterial chromosomes, as well as new computational tools to extract single-molecule molecular signatures from live cells to monitor stoichiometry, spatial localization and mobility in living cells. We also discuss recent developments from our lab made *in vitro*, complementing these *in vivo* studies, which combine optical and magnetic tweezers to manipulate and image single molecules of DNA, with and without bound protein, in a new superresolution fluorescence microscope.

## Introduction

Protein interactions have roles in every facet of the role of DNA in life. They are involved in replicating DNA, proof-reading it and correcting any errors. They package DNA into chromosomal structures and repair it when damaged. They are involved in transcription but also, its regulation. Understanding these interactions is therefore vital to understanding normal cell development but also disease.

Classically, DNA-protein interactions have been probed biochemically (for a review, see Cai and Huang 2012[1]). Electrophoresis is commonly used to detect proteins interacting with DNA in the Electrophoretic mobility shift assay (EMSA). [2] It is also possible to isolate interacting DNA and protein using the Chromatin Immuno-precipitation (ChIP) assay.[3] In this assay bound proteins are chemically cross-linked to DNA in living cells and then chromatin is extracted. DNA-protein complexes can then be isolated by pulling out via suitable antibodies. These types of bulk biochemical assay can only detect average, ensemble dynamics and fail to detect the wide heterogeneity of behaviour present in real molecular systems in biology. Single-molecule techniques can characterize this heterogeneity.

Fluorescence microscopy is the least invasive technique sensitive enough to detect single molecules inside a living cell.[4] It has been used extensively to observe DNA-interacting proteins (see Xie et al. 2008[5] for a review), including the production of individual mRNA molecules in real time in *Escherichia coli*,[6] transcription through the specific binding of *lac* repressor molecules to DNA[7] and DNA break repair through homologous recombination.[8] We have developed a method of fluorescence microscopy, known as Slimfield,[9] for observation of single molecules in live cells over a millisecond time scale, which we have used to elucidate the molecular makeup and time-resolved features of the bacterial replisome[10] and the structural maintenance of chromosome (SMC) complexes which remodel bacterial chromosomes.[11]

## In the cell

In Slimfield microscopy, the normal fluorescence excitation field is reduced to encompass only a single cell (Figure 1A). This produces a Gaussian excitation field (~30μm$^2$) with 100-1000 times the laser excitation intensity of standard epifluorescence microscopy. This allows much greater signal intensity relative to normal camera imaging noise and hence facilitates millisecond time scale imaging of single fluorescently-labelled proteins – this time scale is thus fast enough not only to keep up with the diffusional motion present in the cytoplasm of cells, but can also sample fast molecular transitions that occur in the millisecond regime. This technique was applied to the bacterial replisome.[10] Here, individual *E. coli* replisome components were labelled with the YPet fluorescent protein and expressed from their endogenous promoters, thus resulting in roughly physiological levels of expression. Figure 1B shows fluorescence micrographs of YPet labelled replisome components overlaid on the corresponding brightfield image of the cell. In most cells (75%) two distinct bright spots were observed but in some, replisomes had not moved sufficiently far apart from the origin of replication so a single spot was observed. By analysing the intensity of these spots as they photobleach (Figure 1C), it was possible to quantify the number of YPet in each spot. Repeating this with different YPet labelled replisome components allowed the stoichiometry to be determined.

The replisome is a molecular machine that replicates leading-strand template DNA continuously and lagging-strand template discontinuously (Okazaki fragments) and is made up of more than 11 proteins.[12,13] DnaB is the helicase which separates the strands. Primase binds to the helicase during cycles of priming on the lagging strand. Pol III polymerises the DNA and is held on by the sliding clamp, β, which contains several subunits including τ and δ. Single stranded DNA is bound by single stranded binding protein (Ssb) tetramers which remove secondary structure and protect against nucleases. Historically, it has been accepted that there are two polymerases per replisome[14] but Reyes-Lamonthe et al.[10] found evidence for three Pol III per replisome, associated with three τ units which trimerize the polymerase. Most replisome components formed tight spots in the cell but Ssb had a broader spatial distribution with 8-11 tetramers per replisome.

Slimfield microscopy can also be used in dual-colour fluorescence microscopy and has been used to study the bacterial SMC proteins in *E. coli*.[11] SMC proteins have conserved architecture and function across all domains of life with bacteria using a distant relative called MukB with accessory MukE and MukF proteins playing a role in chromosome segregation and organization.[15,16] Structural and biochemical studies have shown two stoichiometries for the MukBEF complex of 2:4:2 and 2:2:1 (MukB:E:F) dependent on whether ATP is bound or unbound.[17] Dual-colour Slimfield imaging on pairs of GFP and mCherry labelled Muk B, E and F proteins showed that the minimal functional unit had a stoichiometry of 4:4:2 (Figure 2A). MukBEF was found to accumulate in cells in one to three spots of 8-10 of these complexes (Figure 2B) but also in freely diffusing complexes. Fluorescence recovery after photobleaching (FRAP) showed MukBEF in spots exchanged with freely diffusing complexes at a rate of one every 50s and that this was dependent on ATP hydrolysis as no exchange was observed in hydrolysis impaired mutants. Diffusing complexes in both the wild type and hydrolysis mutant were found in the 2:4:2 and 2:2:1 states but in a mutant incapable of nucleotide binding, they were exclusively in 2:4:2. These data point to a 'rock climber' model where MukBEF undergoes multiple cycles of hydrolysis leading to binding and unbinding of DNA, similar to a molecular motor (like kinesin) binding and unbinding its track as it moves. This means MukBEF can capture new DNA segments without releasing from the chromosome, perhaps leading to DNA remodelling.

## Inside a computer

The data obtained from Slimfield microscopy requires a significant amount of *in silico* analysis in order to measure the stoichiometry of single molecular complexes and their motilities. Complexes of fluorescently tagged molecules must be identified by software, the intensity of these spots quantified, their stoichiometry calculated from the photobleaching intensity trace, their position tracked over time to produce a trajectory and the motion in the trajectory characterised.[18] Custom Matlab™ software has been written to perform these analyses and is outlined here.

Before analysing fluorescent signals from inside the cell, it is first necessary to determine the cell boundary. This can be done from standard brightfield images of the cell taken before or after a fluorescence acquisition. These images can be easily segmented if the cell is slightly out of focus, resulting in a black band around the cell. This band can be identified by thresholding the pixel intensity in the image (Figure 2A). These data can be used as a cell mask to identify where in the fluorescence image to look for bright spots corresponding to individual molecular complexes (such as those shown in Figure 1 and 2). Bright spots in the image are identified through a series of image transformations and thresholding steps. These include a thresholded top-hat transformation which pulls out bright signal above the background and then a dilation and expansion to remove individual 'hot' pixels before ultimate erosion leaving a series of candidate spot co-ordinates. If these spots meet criteria including their signal-to-noise ratio, they are accepted by the software. Their centroid is found to sub-pixel precision by iterative Gaussian masking.[19–21] This mask is also used to quantify the background intensity at each spot. This kind of fluorescence data has a significant amount of noise, partially from sources inherent to the instrument but also the varyingly auto-fluorescent environment inside the cell. To circumvent this, spot intensity is quantified as the background-subtracted, total intensity inside the mask.

The spot intensity is plotted over time, decaying as the fluorophores in the complex photobleach. Examples of photobleach traces of YPet labelled replisome components[10] are shown in Figure 1C. The bleaching is exponential and step-like, with the size of the distinct step events being, within experimental noise, an integer multiple of the brightness of a single YPet dye. This unitary fluorescent protein step size in intensity can be quantified and compared against the initial unbleached brightness of the complex to obtain the number of fluorescently labelled molecules present.[22] With multiple spectrally compatible fluorescent tags on molecules in a complex, their stoichiometries can be calculated in the same spot. Photobleach traces can first be filtered using an edge-preserving filter such as a Chung-Kennedy filter,[23] shown in red in Figure 1C, which we had developed from earlier studies for the detection of molecular scale mechanical steps in single-molecule optical tweezers stretch experiments of muscle proteins titin and kettin.[24,25] Pairwise differences between all the intensities in the trace are calculated to produce the Pairwise Difference Distribution Function (PDDF). Fourier spectral analysis of the PDDF produces the power spectrum (Figure 2C, inset) and the fundamental peak of this gives the unitary step size. The exponential photobleach traces are divided by the step size to produce the number of fluorophores active at each time point. The initial number is the number of fluorescently tagged molecules in the spot. For dual-colour investigations a similar method can be applied to the separate colour channels, but in addition utilising co-localization metrics across the cellular images to determine if spots detected in both channels are indeed co-localized over the same regimes of space and time.[26]

The centroid position of each spot at each point in time can be linked together into a trajectory, describing the motion of the complex. We found that MukBEF complexes, when not bound to DNA, freely diffuse in the cytoplasm but exhibit confinement in chromosomally-bound states. There are many modes of motion for *in vivo* molecular complexes. Figure 2C shows simulated data[27] for four common types of diffusive motion. These types of motion can be characterized by their mean square displacement (MSD)[28] as a function of time interval, $\tau$ (Figure 2D). The blue trajectory defines normal diffusion or standard Brownian motion with a linear MSD plot. The cyan plot shows directed motion, such as a molecular motor undergoes moving on a track with upward parabolic MSD plot. An asymptote at high $\tau$, signifies confined diffusion shown in the red plot. Anomalous or sub diffusion (green) has an MSD proportional to $\tau^\alpha$ where $\alpha$ is between 0 and 1 and corresponds to movement through a disordered media. The MSD vs $\tau$ plot of a diffusing complex can be used to characterise its motion[29] but for fluorescence single-molecule data, available time points are limited by photobleaching, resulting in often highly truncated tracks with very little high time interval (τ) data available[30]. More robust characterization of diffusing fluorescent proteins is obtained using Bayesian inference through the 'Bayesian ranking of diffusion' (BARD) method.[27]  This method uses propagators directly to rank normalized posteriors as different modes of diffusion, for example Brownian, directed, anomalous or confined diffusive motion, but which can be generalized to any diffusive mode.

## In a test tube

Although probing protein-DNA interactions is best done in the native cellular environment, *in vitro* experiments allow for direct manipulations of single molecules - not possible inside the cell. *In vitro,* the level of noise is much lower and there is greater scope for using brighter organic dyes rather than dim fluorescent proteins – vastly increasing the effective signal-to-noise ratio for detection. We are developing a microscope which combines optical and magnetic tweezers for manipulating single molecules, with superresolution fluorescence imaging capability.

DNA is helical and its twist is regularly manipulated by proteins: helicases unwrap it as it is replicated, topoisomerases regulate its coiling and it is wrapped up and packaged by chromatins in the nucleus. Thus, for many protein-DNA interactions, manipulating and measuring the forces and torques involved is key to understanding the dynamics involved. Magnetic tweezers are a powerful physical science tool to achieve this, since torque can be applied to single DNA molecules via a suitable magnetic probe following controlled rotation of the B-field inside the magnetic tweezers sample chamber. The canonical design uses DNA tethered at one end to a coverslip surface in a microscope with the other attached to a ferromagnetic or paramagnetic bead  so that the helical axis of the tethered DNA is perpendicular to the field of view.[31] In its simplest design, a permanent magnet is typically mounted above the sample chamber, allowing torque to be applied to the magnetic bead, which is chemically conjugated to the DNA via multiple torsionally-constrained bonds that prevent free rotation of the bead relative to the DNA. This approach does not permit easy adjustment of the amount of torque or independent measurement and application of force and torque, and so more complicated designs using electromagnets and multiple coils around inverted microscopes have been built.[32,33]

Magnetic tweezers alone struggle to manipulate DNA torque and end-to-end extension independently in 3D; for this, optical tweezers represent a promising tool.[34] Systems with both optical and magnetic tweezers have been built before[35,36] but the movement of the proteins and DNA investigated with such systems was inferred from bead position rather than the molecules directly. Fluorescence microscopy can be incorporated to directly image the molecules involved, both the DNA and bound/translocating proteins, and has been integrated in the canonical magnetic tweezers set up[37,38] but with DNA perpendicular to the field of view so it cannot be imaged along its length.  Transverse magnetic tweezers in principle allow fluorescently-labelled DNA to be visualized in the focal plane of a fluorescence microscope whilst applying different levels of controlled twist to the DNA.[39] We have designed a microscope which combines transverse magnetic and optical tweezers, as well as superresolution fluorescence microscopy. The design schematic is shown in

Figure 3A, with two coaxial, parallel pairs of electromagnetic coils arranged perpendicularly, allowing the optical trapping laser beam to pass through unimpeded.

Multiple methods exist to circumvent the diffraction limit and obtain superresolution fluorescence images but they all rely on switching fluorophores in a densely labelled sample such that only a sub-population are 'on' (i.e. photoactive) at any particular sampling time point. This allows for precise determination of each fluorophore's position using the iterative fitting/masking algorithms discussed previously which are then used to reconstruct a superresolution image. Different methods for achieving this use photoactivatable fluorophores in PALM[40] and photoblinking in STORM[41], binding and unbinding of fluorophores in BaLM[42] and BaLM photobleaching [43]. These techniques have been used to image DNA using fluorescent chemical dyes which bind to different parts of the DNA. Intercalating dyes such as YOYO have been used for super resolution imaging,[44] as well as minor groove binding dyes, which potentially perturb the native DNA structure less, such as SYTO-13[45] and PicoGreen.[46] Covalently bound dyes have also been used with much higher coverage.[47] Modifying the imaging pathway of the microscope with the addition of a cylindrical lens results in controllable astigmatism of fluorescence images, which can then be used as a metric for how far above or below the focal plane a given detected dye molecule is; this therefore facilitates 3D determination of spatial localization of the dye, with a typical precision of ~30-40 nm laterally and ~50-60 nm axially when sampling at video-rate time resolution.[18]

We have developed superresolution methods for imaging DNA. Using λ-DNA as a control sample firmly immobilized to a glass coverslip, we have generated, as proof-of-principle in our new device, superresolution images using the photobleaching and blinking of YOYO dye bound to DNA (Figure3B). Bright spots were found using the same bespoke automated algorithms as before, and their positions determined using the same methods used for our *in vivo* experiments, confirming superresolution images DNA with 30-40nm lateral resolution.

## Summary and outlook

Observing protein-DNA interactions at the single-molecule level allows the full heterogeneity of complex molecular behaviour to be characterized. We have developed Slimfield microscopy which allows us to observe single molecules of fluorescently tagged proteins in living cells at exceptionally high millisecond time resolution. With careful *in silico* analysis of these data using bespoke computational tools, the stoichiometry of the observed molecular complexes can be determined. This was used to uncover the dynamics and architecture of the bacterial replisome, which bring about faithful DNA replication, and of SMC proteins, which remodel DNA. *In vitro* experiments allow for increased signal-to-noise detection ratios, and for topological manipulation of single molecules of DNA which cannot be done in live-cell experiments. We have designed a new technology which combines optical and magnetic tweezers as well as superresolution fluorescence imaging for mechanical investigations of single molecules of DNA. By combining full molecular manipulation and measurement of the forces and torques crucial to DNA-interacting proteins with superresolution imaging, we hope to gain unprecedented insight into the mode of action of proteins which operate through binding to DNA, as well as of the intrinsic dynamic topological properties of DNA molecules themselves which are essential to their biological function. Perhaps more importantly this is a prime example of the emergence of a suite of new, innovative physical science tools at the molecular level,[48] which one might argue constitutes a subset of a 'toolbox', which can be called upon to address focused, unresolved questions from the life sciences.

## Acknowledgements

We thank Christoph Baumann (University of York) for useful discussion of DNA *in vitro* experiments. Many thanks to David Sherratt and his team (University of Oxford) for collaborating on the DNA replisome and remodelling work.

## Funding

Our work is supported by the Biological Physical Sciences Institute (BPSI), University of York.

## References

[1]     Cai Y-H, Huang H. (2012) Advances in the study of protein-DNA interaction. Amino Acids. **43**, 1141–6.

[2]     Hellman LM, Fried MG. (2007) Electrophoretic mobility shift assay (EMSA) for detecting protein-nucleic acid interactions. Nat. Protoc. **2**, 1849–61.

[3]     Furey TS. (2012) ChIP-seq and beyond: new and improved methodologies to detect and characterize protein-DNA interactions. Nat. Rev. Genet. **13**, 840–52.

[4]     Leake M. (2013) The physics of life: one molecule at a time. Philos. Trans. R. ….

[5]     Xie XS, Choi PJ, Li G-W, Lee NK, Lia G. (2008) Single-molecule approach to molecular biology in living bacterial cells. Annu. Rev. Biophys. **37**, 417–44.

[6]     Golding I, Paulsson J, Zawilski SM, Cox EC. (2005) Real-time kinetics of gene activity in individual bacteria. Cell. **123**, 1025–36.

[7]     Elf J, Li G-W, Xie XS. (2007) Probing transcription factor dynamics at the single-molecule level in a living cell. Science. **316**, 1191–4.

[8]     Lesterlin C, Ball G, Schermelleh L, Sherratt DJ. (2014) RecA bundles mediate homology pairing between distant sisters during DNA break repair. Nature. **506**, 249–53.

[9]     Plank M, Wadhams GH, Leake MC. (2009) Millisecond timescale slimfield imaging and automated quantification of single fluorescent protein molecules for use in probing complex biological processes. Integr. Biol. (Camb). **1**, 602–12.

[10]    Reyes-Lamothe R, Sherratt DJ, Leake MC. (2010) Stoichiometry and architecture of active DNA replication machinery in Escherichia coli. Science. **328**, 498–501.

[11] Badrinarayanan A, Reyes-Lamothe R, Uphoff S, Leake MC, Sherratt DJ. (2012) In vivo architecture and action of bacterial structural maintenance of chromosome proteins. Science. **338**, 528–31.

[12] McInerney P, Johnson A, Katz F, O'Donnell M. (2007) Characterization of a triple DNA polymerase replisome. Mol. Cell. **27**, 527–38.

[13] Johnson A, O'Donnell M. (2005) Cellular DNA replicases: components and dynamics at the replication fork. Annu. Rev. Biochem. **74**, 283–315.

[14] Alberts BM, Barry J, Bedinger P, Formosa T, Jongeneel CV, Kreuzer KN. (1983) Studies on DNA Replication in the Bacteriophage T4 a--.gif System. Cold Spring Harb. Symp. Quant. Biol. **47**, 655–668.

[15] Niki H, Imamura R, Kitaoka M, Yamanaka K, Ogura T, Hiraga S. (1992) E.coli MukB protein involved in chromosome partition forms a homodimer with a rod-and-hinge structure having DNA binding and ATP/GTP binding activities. EMBO J. **11**, 5101–9.

[16] Yamanaka K, Ogura T, Niki H, Hiraga S. (1996) Identification of two new genes, mukE and mukF, involved in chromosome partitioning in Escherichia coli. Mol. Gen. Genet. **250**, 241–51.

[17] Woo J-S, Lim J-H, Shin H-C, Suh M-K, Ku B, Lee K-H, Joo K, Robinson H, Lee J, Park S-Y, Ha N-C, et al. (2009) Structural studies of a bacterial condensin complex reveal ATP-dependent disruption of intersubunit interactions. Cell. **136**, 85–96.

[18] Leake MC. (2014) Analytical tools for single-molecule fluorescence imaging in cellulo. Phys. Chem. Chem. Phys. **16**, 12635–47.

[19] Thompson RE, Larson DR, Webb WW. (2002) Precise nanometer localization analysis for individual fluorescent probes. Biophys. J. **82**, 2775–83.

[20] Leake MC, Greene NP, Godun RM, Granjon T, Buchanan G, Chen S, Berry RM, Palmer T, Berks BC. (2008) Variable stoichiometry of the TatA component of the twin-arginine protein transport system observed by in vivo single-molecule imaging. Proc. Natl. Acad. Sci. U. S. A. **105**, 15376–81.

[21] Delalez NJ, Wadhams GH, Rosser G, Xue Q, Brown MT, Dobbie IM, Berry RM, Leake MC, Armitage JP. (2010) Signal-dependent turnover of the bacterial flagellar switch protein FliM. Proc. Natl. Acad. Sci. U. S. A. **107**, 11347–51.

[22] Leake MC, Chandler JH, Wadhams GH, Bai F, Berry RM, Armitage JP. (2006) Stoichiometry and turnover in single, functioning membrane protein complexes. Nature. **443**, 355–8.

[23] Chung SH, Kennedy RA. (1991) Forward-backward non-linear filtering technique for extracting small biological signals from noise. J. Neurosci. Methods. **40**, 71–86.

[24] Leake MC, Wilson D, Bullard B, Simmons RM. (2003) The elasticity of single kettin molecules using a two-bead laser-tweezers assay. FEBS Lett. **535**, 55–60.

[25] Leake MC, Wilson D, Gautel M, Simmons RM. (2004) The elasticity of single titin molecules using a two-bead optical tweezers assay. Biophys. J. **87**, 1112–35.

[26] Llorente-Garcia I, Lenn T, Erhardt H, Harriman OL, Liu L-N, Robson A, Chiu S-W, Matthews S, Willis NJ, Bray CD, Lee S-H, et al. (2014) Single-molecule in vivo imaging of bacterial respiratory complexes indicates delocalized oxidative phosphorylation. Biochim. Biophys. Acta. **1837**, 811–24.

[27] Robson A, Burrage K, Leake MC. (2013) Inferring diffusion in single live cells at the single-molecule level. Philos. Trans. R. Soc. Lond. B. Biol. Sci. **368**, 20120029.

[28] Qian H, Sheetz MP, Elson EL. (1991) Single particle tracking. Analysis of diffusion and flow in two-dimensional systems. Biophys. J. **60**, 910–21.

[29] Kusumi A, Sako Y, Yamamoto M. (1993) Confined lateral diffusion of membrane receptors as studied by single particle tracking (nanovid microscopy). Effects of calcium-induced differentiation in cultured epithelial cells. Biophys. J. **65**, 2021–40.

[30] Füreder-Kitzmüller E, Hesse J, Ebner A, Gruber HJ, Schütz GJ. (2005) Non-exponential bleaching of single bioconjugated Cy5 molecules. Chem. Phys. Lett. **404**, 13–18.

[31] De Vlaminck I, Dekker C. (2012) Recent advances in magnetic tweezers. Annu. Rev. Biophys. **41**, 453–72.

[32] Claudet C, Bednar J. (2005) Magneto-optical tweezers built around an inverted microscope. Appl. Opt. **44**, 3454.

[33] Janssen XJA, Lipfert J, Jager T, Daudey R, Beekman J, Dekker NH. (2012) Electromagnetic torque tweezers: a versatile approach for measurement of single-molecule twist and torque. Nano Lett. **12**, 3634–9.

[34] Moffitt JR, Chemla YR, Smith SB, Bustamante C. (2008) Recent advances in optical tweezers. Annu. Rev. Biochem. **77**, 205–28.

[35] Crut A, Koster DA, Seidel R, Wiggins CH, Dekker NH. (2007) Fast dynamics of supercoiled DNA revealed by single-molecule experiments. Proc. Natl. Acad. Sci. U. S. A. **104**, 11957–62.

[36] Van Loenhout MTJ, De Vlaminck I, Flebus B, den Blanken JF, Zweifel LP, Hooning KM, Kerssemakers JWJ, Dekker C. (2013) Scanning a DNA molecule for bound proteins using hybrid magnetic and optical tweezers. PLoS One. **8**, e65329.

[37] Shroff H, Reinhard BM, Siu M, Agarwal H, Spakowitz A, Liphardt J. (2005) Biocompatible Force Sensor with Optical Readout and Dimensions of 6 nm 3. Nano Lett. **5**, 1509–1514.

[38] Hugel T, Michaelis J, Hetherington CL, Jardine PJ, Grimes S, Walter JM, Falk W, Anderson DL, Bustamante C. (2007) Experimental test of connector rotation during DNA packaging into bacteriophage phi29 capsids. PLoS Biol. **5**, e59.

[39] Graham JS, Johnson RC, Marko JF. (2011) Concentration-dependent exchange accelerates turnover of proteins bound to double-stranded DNA. Nucleic Acids Res. **39**, 2249–59.

[40] Hess ST, Girirajan TPK, Mason MD. (2006) Ultra-high resolution imaging by fluorescence photoactivation localization microscopy. Biophys. J. **91**, 4258–72.

[41] Rust MJ, Bates M, Zhuang X. (2006) Sub-diffraction-limit imaging by stochastic optical reconstruction microscopy (STORM). Nat. Methods. **3**, 793–5.

[42] Schoen I, Ries J, Klotzsch E, Ewers H, Vogel V. (2011) Binding-activated localization microscopy of DNA structures. Nano Lett. **11**, 4008–11.

[43] Burnette DT, Sengupta P, Dai Y, Lippincott-Schwartz J, Kachar B. (2011) Bleaching/blinking assisted localization microscopy for superresolution imaging using standard fluorescent molecules. Proc. Natl. Acad. Sci. U. S. A. **108**, 21081–6.

[44] Flors C. (2011) DNA and chromatin imaging with super-resolution fluorescence microscopy based on single-molecule localization. Biopolymers. **95**, 290–7.

[45] Flors C, Earnshaw WC. (2011) Super-resolution fluorescence microscopy as a tool to study the nanoscale organization of chromosomes. Curr. Opin. Chem. Biol. **15**, 838–44.

[46] Benke A, Manley S. (2012) Live-cell dSTORM of cellular DNA based on direct DNA labeling. Chembiochem. **13**, 298–301.

[47] Zessin PJM, Finan K, Heilemann M. (2012) Super-resolution fluorescence imaging of chromosomal DNA. J. Struct. Biol. **177**, 344–8.

[48] Leake MC. (2013) The physics of life: one molecule at a time. Philos. Trans. R. Soc. Lond. B. Biol. Sci. **368**, 20120248.

# Figures

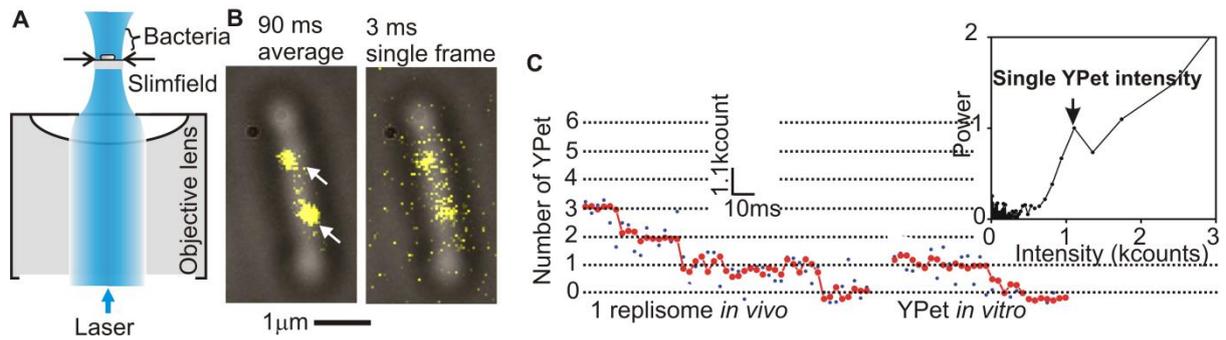

**Figure 1:** Slimfield microscopy and photobleach analysis of the *E. coli* replisome. **A** Schematic of Slimfield microscopy. An intense Gaussian field encompassing a single *E. coli* cell is generated at the sample. **B** Overlaid brightfield (grey) and 90 ms frame-averaged fluorescence images (yellow) of replisome polymerase labelled strain, bright spots marked with arrows, corresponding single 3 ms frames. **C** Raw intensity (blue) and Chung-Kennedy filtered data (red) for a single replisome spot, marked here with DnaQ-YPet compared against a single molecule of surface-immobilized YPet *in vitro*, with (inset) Fourier spectral analysis for a photobleach trace indicating brightness of a single YPet.

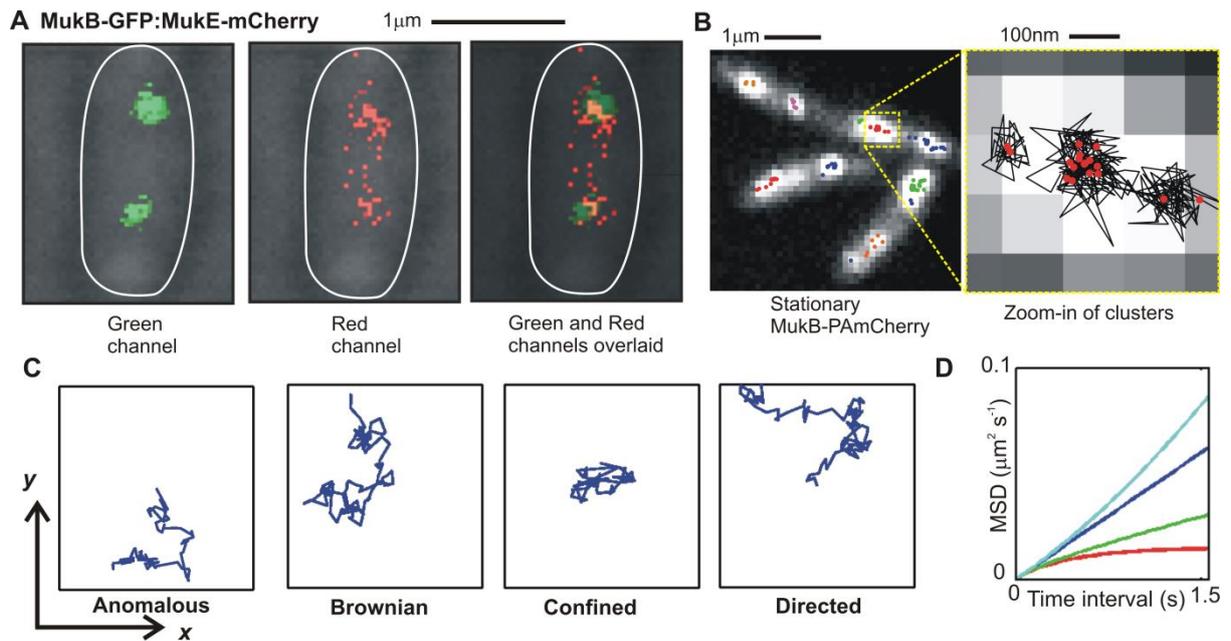

**Figure 2:** Co-localization and diffusion analysis. **A** Overlayed brightfield (gray), segmented to show cell outline, and dual colour fluorescence images, MukB-GFP in green and MukE-mCherry in red. **B** Fluorescence micrographs of MukB-PAmCherry (greyscale) with trajectories overlayed in colour, here showing relatively immobile spots. **C** Simulated diffusive trajectories in lateral *xy* focal plane with **D** corresponding MSD vs τ plot on the right. Different diffusive modes - anomalous, Brownian, confined and directed – shown in green, blue, red and cyan respectively.

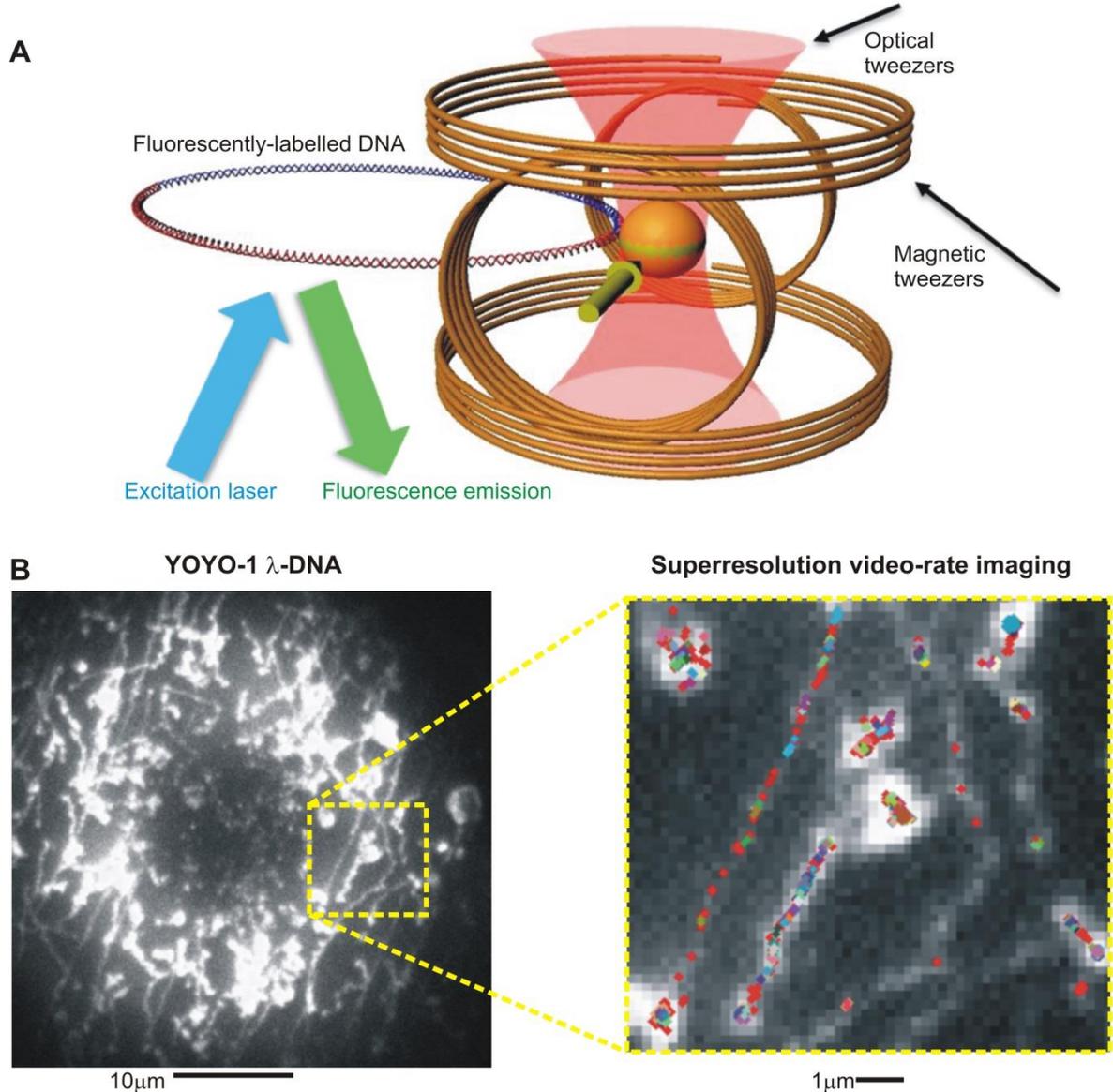

**Figure 3: A** Schematic of our magnetic and optical tweezers design. Two pairs of electromagnetic coils are arranged perpendicularly in so-called Helmholtz configurations, allow the optical trapping laser beam to pass through unimpeded and the circular DNA construct to be imaged simultaneously using fluorescence excitation. **B** Fluorescence micrograph of YOYO-1 labelled DNA (left panel) with zoomed-in section overlaid with superresolution reconstruction fits (right panel).